\documentstyle[12pt,aaspp4]{article}

\newcommand{\hda}{ H(z) D(z)}
\newcommand{\hdaf}{ H_{s}(z) D_{s}(z)}
\newcommand{\rxi}{ \tilde{\xi}({\bf w}) }

\newcommand{\kms}{\, {\rm km\, s}^{-1}}

\newcommand{\mpc}{\, {\rm Mpc}}

\newcommand{\etal}{et al.\ }

\begin{document}

\title{Geometric Distortion of the Correlation function of Lyman-break Galaxies}
\author{Vibhat Nair}
\affil{ University of Pennsylvania, Department of Physics and Astronomy, 
Philadelphia, PA 19104}
\authoremail{ nair@student.physics.upenn.edu}

\begin{abstract}
The number of galaxies with measured redshifts $z \gtrsim 1$ is at present 
rapidly increasing, allowing for measurements of their correlation function.
The correlations function $\xi ( \psi,v)$ is measured in redshift space, as a 
function of the angular separation $\psi $ and velocity difference $v$. The relation 
between angle and velocity difference depends on the cosmological model through
the factor $H(z) \cdot D(z)$, where $H(z)$ is the Hubble parameter and $D(z)$ is 
the angular diameter distance. Therefore, the cosmological model can be constrained
by measuring this factor from the shape of the contours of the $\xi(\psi,v)$, 
if the effect of peculiar velocities can be taken into account. Here, we 
investigate this method applied to the high redshift Lyman-break galaxies. 
The high bias factor of this galaxy population should suppress peculiar 
velocity effects, leaving the cosmological distortion as the main contribution
to the anisotropy of the correlation function. We estimate the shot noise and 
cosmic variance errors using linear theory. A field size of at least $0.2 \, 
{\rm deg}^{2}$
is required to distinguish the Einstein-de Sitter model from the flat 
$\Lambda=0.7$ model, if $1.25$
Lyman-break galaxies are measured per square 
arc minute. With a field of $1 \, {\rm deg}^{2}$, the cosmological constant can 
be measured to $\sim 20 \%$ 
accuracy if it is large ($\gtrsim 0.5$). Other equations of state for a scalar 
field can also be constrained.

\end{abstract}

\keywords{cosmology:theory - large-scale structure of Universe}

\section{Introduction}

  Alcock \& Paczy\'nski (1979) suggested the possibility of using the
clustering statistics of galaxies in redshift space to constrain the
global geometry in the universe. The basic idea is that, since clusters
of galaxies should not be preferentially aligned along any direction
relative to a fixed observer, their average shape ought to be
spherically symmetric. Therefore, if galaxies were following the
Hubble expansion of the universe, without any peculiar velocities,
 the average extent of clusters in radial velocity $v_r$
(measured from redshifts) and their angular size $\psi$ 
are related to the physical size of the cluster $L$ by $v_r = H(z)\, L$,
and $\psi = L/D(z)$, respectively. Here, $H(z)$ and $D(z)$ are the
Hubble constant and the angular diameter distance at the redshift $z$
where the clusters are observed. The condition that clusters are spherical
on average can then yield the value of $H(z)\cdot D(z)$. Of course, the effect of 
peculiar velocities must be included in
order to apply this method, since any clustering induced by gravity
will generally introduce peculiar velocities (Kaiser 1987) that will
cause a distortion of similar or greater magnitude than the differences
between cosmological models.

Recently, the rate at which galaxies at high redshift are being
identified has dramatically increased thanks to the Lyman limit
technique, using the fact that the reddest objects among faint galaxies
will often be galaxies at the redshift where the
Lyman limit wavelength is between the two bands used to measure the 
color (Guhathakurta \etal 1990; Steidel \& Hamilton 1993; Steidel \etal 1996).
For example, very red objects in $U-B$ are likely to be
galaxies at redshift $z\simeq 3$.

The galaxy correlation function, $\xi({\bf r})$, which measures the probability in excess
of a random distribution of finding a galaxy at a real space separation vector ${\bf r}$ from 
another galaxy,
has been measured for the first time for the population of Lyman-break galaxies (Giavalisco \etal 1998). The correlation length, 
defined to be the separation at which the excess probability is equal to that of a 
random distribution, has been estimated to be $\sim 2.1 h^{-1}$ Mpc 
(for an $\Omega_{0}=1$ universe;  the symbol $\Omega$ is used here for the ratio
of the density of matter in the universe to the critical density, the subscript $0$ indicates 
redshift zero), about half of the 
correlation length of galaxies at $z=0$.
The bias, defined as the ratio of the correlation function of galaxies to that of matter
at a fixed separation, is estimated to be large, $\sim 4$ for an $\Omega_{0}=1$ universe and 
smaller for universes with smaller dark matter content (Giavalisco \etal 1998).
Count-in-cells analysis of the Lyman-break sample used in conjunction with a Press-Schechter
mass function for the halos also indicate that these galaxies are 
likely to reside in rare, massive halos that existed at the time (Adelberger \etal 1998; Steidel \etal 1998; see also Coles \etal 1998 and Wechsler \etal 1998 for models of clustering of Lyman-break galaxies). These rare halos are 
expected to be much more clustered than the underlying matter distribution as originally 
suggested by Kaiser (1984) (see also Mo \& white 1996, for analytic models of bias as a 
function of the mass of halos). 
Both these analysis indicate that 
the population of Lyman-break galaxies is likely to be  highly biased with respect to the 
underlying matter distribution.  


In this paper we investigate the 
feasibility of using the distortion of the redshift space correlation function 
of this population of galaxies to measure cosmological parameters. 
This possibility has been suggested before by Matsubara and 
Suto (1996) who proposed using the ratio of the value of the correlation 
function 
parallel to the line of sight to its value perpendicular to the line 
of sight at a fixed separation as a measure of the distortion. 
In this paper we express the angular 
dependence of the cosmological redshift space distortion of the correlation function 
as a multipole expansion. 
We are also specifically interested in applying this method 
to the highly biased, 
high redshift population of Lyman-break galaxies.  Ballinger 
\etal (1996) have investigated the use of the full
functional 
form of the redshift space power spectrum to separately measure the 
peculiar velocity effects and cosmological geometry effects.
In essence this reduces to  using both
the quadrupolar as well as the octapolar distortion of the redshift space 
power spectrum to simultaneously constrain 
the cosmological constant as well as the parameter
$\beta=\Omega^{0.6}/b$, where b is the linear theory bias.
In this paper we fix the bias of the galaxy distribution by using the 
constraints on the matter power spectrum at redshift zero derived from 
observations of cluster abundances. On large scales the power spectrum of 
matter at any redshift is related to the power spectrum at redshift zero 
through the linear growth factor. 
We can then use the lowest order quadrupolar distortion of the power
spectrum 
alone to constrain other cosmological parameters such as the cosmological
constant.
Ballinger \etal (1996) also  estimated the errors involved in such a survey although in 
Fourier space. We estimate the errors 
in estimating cosmological parameters directly from the correlation function. 

  On sufficiently large scales, where density fluctuations are in the
linear regime, the angular form of the redshift space correlation
function depends only on two parameters: the cosmological term
$H(z)\cdot D(z)$, and the bias of the galaxy population.
This paper presents a general method of estimating these two
parameters from the basic data of a galaxy redshift survey, and
evaluates the size of the survey that is necessary to determine the
two parameters (or a combination of them, given other constraints from
the galaxy distribution at the present time) with a given accuracy.
We shall analyze the sensitivity of the method to a variety of
cosmological models, placing special emphasis on models that
contain a cosmological constant or a new component of the energy
density of the universe with negative pressure christened Quintessence 
(e.g. Kodama \& Sasaki 1984; Peebles \& Ratra 1988;  Caldwell \etal 1998), 
given the recent evidence from the luminosity distances to Type Ia
supernovae (Garnavich \etal 1998; Perlmutter \etal 1997; Reiss \etal
1998)  suggesting an accelerating universe. As pointed out by
Alcock \& Paczy\'nski, the quantity $H(z)\cdot D(z)$ is more
sensitive to this type of component than to space curvature.

  The paper is arranged as follows. In \S 2 we describe the effect
of geometric distortion. In \S 3 we introduce the method for 
measuring the effects of cosmological geometry and peculiar velocity
effects on the redshift space correlation function.
In \S 4 we present predictions for a variety of cosmological
models, and in \S 5 we estimate the errors in the observational
determination of the redshift space correlation function contributed
by shot noise and by the finite size of the observed volume. Our
discussion is given in \S 6.

\section{Method}

  A redshift survey consists of measuring the radial velocity and
angular position of every galaxy included in the sample. We denote by
${\bf n}$ the unit vector along the line of sight, which, if the survey
does not extend over a very large area, can be considered constant for
all galaxies. Given a
pair of galaxies, let $v$ be the difference between their radial
velocities, and $\psi$ their angular separation. We define their
vector separation in redshift space ${\bf w}$ as (see Figure 1)
\begin{eqnarray}
{\bf w} \cdot {\bf n} & = & v ~ , \nonumber  \\
| {\bf w} - ({\bf w} \cdot {\bf n}) {\bf n} | &
= & H(z)D(z) \, \psi ~ , \nonumber \\
w^2 & = & v^2 +
\left[  H(z)D(z)\, \psi  \right]^2 ~ .
\end{eqnarray}
where $H(z)$ and $D(z)$ are the Hubble constant and the angular diameter 
distance at the mean redshift of the survey, $z$. 
We also define $\mu$, for future use, as the cosine of the angle between
the vector separation between two galaxies and the line of sight:
\begin{equation}
\mu=\frac{v}{w}
\end{equation}

The quantity
$H(z)D(z)$ contains the dependence on the cosmological model. If we
could measure the correlation function of galaxies directly in real
space (measuring distances to galaxies instead of radial velocities),
then the simple requirement that the correlation function should be
isotropic would yield the value of $H(z)D(z)$. However, peculiar
velocities should obviously introduce an anisotropy in the correlation
function, and their effect needs to be included.

\subsection{Model Dependence of $\hda$}
Figures 2 and 3 show the  ratio $\hda / \hdaf$ for various models, where
$\hdaf$ is the value of $\hda$ for a ``fiducial'' model, here adopted
to be the Einstein-de Sitter model, with $\Omega_0=1$ in the form of
pressureless matter. The symbol $\Omega_0$ is used here for the present ratio
of the density of matter in the universe to the critical density.

Two of the  models
shown in Figures 2 and 3 are the open model (with space curvature but no
negative pressure components) and the cosmological constant (or $\Lambda$)
model (with no space curvature and a component with pressure
$p=-\rho c^2$). The third of the models shown is a Quintessence or Q model with no 
spatial curvature and a 
component with equation of state $p=-\rho c^2/3$. 

The quantity $\hda$ is much more sensitive to
$\Lambda$ than to space curvature, and is also 
sensitive to the Q model, with a different redshift dependence. 
In general, a component of the energy density
in the universe with negative pressure can have any equation of state,
but the case $p=-\rho c^2/3$ implies an expansion mimicking exactly
that of an open universe. Therefore, $H(z)$ in our Q model is exactly
the same as in the open model. However, whereas in the open model the
negative space curvature increases the angular diameter distance compared
to the Einstein-de Sitter model, cancelling almost exactly the decrease
in $H(z)$, the flat geometry of the Q model results in smaller angular
diameter distances, so $\hda$ is smaller than in the Einstein-de Sitter
model due to the decrease of $H(z)$.

  It is useful to note at this point that in order to obtain useful
constraints on cosmological models, $\hda$ must be measured to an accuracy
better than $\sim 10 \%$. In order to distinguish, between
a cosmological constant and a Q model, $\hda$ must of course
be measured at several redshifts with even higher accuracy. In practice,
we can expect that any constraints obtained from measuring $\hda$ should
be combined with other knowledge obtained, for example, from the
luminosity distances to Type Ia supernovae.

\section{Effect of peculiar velocities on the redshift space correlation function }

For a given value of $\hda$ the effect of 
peculiar velocities on the shape of the redshift space correlation
function is well described in the literature (e.g.  McGill 1990; Hamilton 1992;Fisher 1995)
and 
the redshift space correlation function, $\tilde{\xi}({\bf w})$,
is given by : 
\begin{eqnarray}
\tilde{\xi}({\bf w})&=&\sum_{l=0,2,4} D_{l} (\beta,w,z) \cdot P_{l}(\mu) \, ,
\label{eqnxi}
\end{eqnarray}
where,
\begin{eqnarray}
\beta \simeq~\frac{\Omega(z)^{0.6}}{b(z)}, \nonumber 
\end{eqnarray}
where $b(z)$ is the bias parameter for the class of objects 
under survey and $\Omega(z)$ is the ratio of the density of matter to the critical density 
at redshift $z$. The coefficients of the expansion in Legendre polynomials, 
$D_{l}$, can be expressed as:
\begin{eqnarray}
D_{l} (\beta,w,z) &=& (-1)^{l} \cdot A_{l} (\beta) \cdot \xi_{l}(w,z) ~,
\end{eqnarray}
where
\begin{eqnarray}
A_{0}&=&\left({1+ \frac{2}{3} \beta + \frac{1}{5} \beta^{2}} \right)~, \nonumber \\
A_{2}&=&\left( {\frac{4}{3} \beta - \frac{4}{7} \beta^{2} } \right)~, \nonumber \\
A_{4}&=& \left( {\frac{8}{35} \beta^{2}} \right)~, \nonumber 
\label{eqnAl}
\end{eqnarray}
and
\begin{eqnarray}
\xi_{l}(w,z)&=& \frac{b(z)^{2}}{2 \pi^2} \int dk~k^{2}~P(k,z)~j_{l}(kw)~, 
\end{eqnarray}
and $j_{l}$ is the lth order spherical Bessel function. The function $P(k,z)$ 
is  the linear matter power spectrum at redshift z in terms of the k vector
in velocity space.

Note that $\xi_{0}(w,z)$ is proportional to the real space matter correlation 
function at redshift z. Hence, $D_{0}$ is equal to the real space two point 
correlation function for this class of objects except for the factor of 
$A_{0}[\beta(z)]$. We also mention here that the $D_{2}$ coefficient is 
negative implying a squashing of the contours of the correlation 
function along the line of sight as is expected due to the peculiar 
velocities from infall on large scales.

Throughout this paper we use the simple model of linear, local, constant  biasing of galaxies, 
i.e.   the overdensity in the number of galaxies, 
$\delta_{g}(\vec{w})$, is given by $b \times 
\delta_{m}(\vec{w})$, where $\delta_{m}(\vec{w})$ is the overdensity in matter and b the bias.
For general deterministic local bias models, this is valid in linear theory where 
$\delta_{g} < 1$ ( for  $b>1$ and 
$\delta_{m} \sim -1, \, \delta_{g} < -1$ is unphysical) (Gazta\~naga \& Baugh 1998). 
Hence, our results are likely to be valid 
on large scales where the correlation function is smaller than one. 
In reality, biasing is not easily modeled since it depends on the complex process of 
galaxy formation, which is poorly understood.
Several alternative 
models of galaxy biasing have been suggested, including non-local biasing mechanisms 
(Babul \& White 1991; Bower \etal 1993) and stochastic biasing 
(Dekel \& Lahav 1998; Tegmark \& Peebles 1998). However, for stochastic (local) models, 
on large scales, the bias (the ratio of the correlation function of galaxies to that of 
matter) will be independent of scale  (Scherrer and Weinberg 1998) as in 
the case of a linear, local, constant biasing scheme, although the variance in the measured 
correlation function will be larger for such models. 
On the other hand, non-local models of galaxy biasing in which the efficiency of galaxy 
formation is modulated coherently over large scales, result in scale dependent bias.
In the absence of a well motivated model for bias, we have assumed the simplest
scale independent model for the bias. It is valid only on 
large scales and is not generally valid for non-local biasing
models. 
We also mention that we have only taken the linear infall velocities into account in 
calculating the redshift space correlation function (see Equation \ref{eqnxi}). On 
small scales non-linear velocity effects (`Fingers of God') will also be important 
(e.g. see  Fisher \etal 1994 for the redshift space correlation function of IRAS galaxies).

\subsection{Effect of geometric distortion}

In order to test the magnitude of the geometric distortion, we calculate the 
anisotropy introduced in the correlation function by varying 
$\hda$ about its fiducial value, $\hdaf$. Let the  product $\hda$ for 
any other model be given by 
\begin{equation}
\hda = \hdaf \cdot \sqrt{1+\alpha(z)}~,
\end{equation}
where $\alpha(z)$ is defined as the geometric distortion parameter.
Then, using equations (1) and (2)  we have,
\begin{eqnarray}
w^{2} &=& w_{s}^{2} ~\eta ^{2} (\alpha,\mu_{s}) ~, \nonumber \\
\mu^{2} &=&\frac{\mu_{s}^{2}}{\eta ^{2}(\alpha,\mu_{s})} ~,
\end{eqnarray}
where 
\begin{equation}
\eta^{2} (\alpha,\mu_{s}) = 1 + \alpha (1 - \mu_{s}^{2}) ~. 
\end{equation}
We can now express equation (\ref{eqnxi}) in terms of the variables 
$w_{s}, \mu_{s}$ in the fiducial model: 
\begin{equation}
\rxi = \sum_{l=0,2,4} D_{l}(\beta,w_{s} \cdot \eta,z) \cdot P_{l}(\frac{\mu_{s}}{\eta})~.
\end{equation}
Rewriting this as a series in $P_{l}(\mu_{s})$,
\begin{equation}
\rxi =\sum_{l} C_{l} (\beta,w_{s},z) \cdot P_{l}(\mu_{s})~,
\end{equation}
one can immediately see from the angular dependence in $\eta$ that the 
expansion is an infinite series in $P_{l}(\mu_{s})$,
with the new coefficients of the Legendre polynomials, $C_{l}$, being given 
by,
\begin{equation}
C_{n}(\beta,w_{s},z)=\left( {\frac{2n+1}{2}} \right) \sum_{l=0,2,4} \int D_{l} (\beta,w_{s} \cdot \eta,z) \cdot P_{l}(\frac{\mu_{s}}{\eta}) \cdot P_{n}(\mu_s)~d\mu_{s}~.
\end{equation}
Thus, expressing the coefficients $D_{l}$ of a given model in terms of 
fiducial coordinates introduces angular distortion in the redshift space 
correlation function.

\section{Results for the geometric distortion}

In this section we  present our results for the sensitivity of the anisotropy
of the correlation function to the geometric distortion parameter, $\alpha$.
We consider here a galaxy survey with a mean redshift of $3$,
the typical redshift of the current Lyman limit galaxy surveys. Our  
fiducial model is $\Omega_{0}=1.0$,
with the standard cold dark matter (SCDM) power spectrum. On large scales the 
power spectrum at redshift $3$ is related to the power spectrum at redshift zero 
by the linear growth factor. We adopt the cluster normalization 
for the power spectrum at redshift zero, obtained by requiring that the 
observed density of galaxy clusters with a given X-ray temperature
matches the theoretical prediction. 
The constraint obtained in this way can be expressed in terms of the 
fluctuation in a sphere of radius $8h^{-1}$ Mpc, $\sigma_{8}$,
given by (Eke \etal 1996):
\begin{eqnarray}
\sigma_{8}&=&0.52~\Omega_{0}^{-0.46 \Omega_{0}}\,,\,{\rm for}\, \Lambda_{0}=0 ~,\nonumber 
\end{eqnarray}
and
\begin{eqnarray}
\sigma_{8}&=&0.52~\Omega_{0}^{-0.52 \Omega_{0}}\,,\,{\rm for}~\Omega_{0}+\Lambda_{0}=1 .
\end{eqnarray}

In Figure 4 we plot the $C_{l}(\beta,w_{s})$ coefficients for $l=0,2$ and
$4$,
for our fiducial model (lighter lines) and the $\Lambda$ model with 
$\Omega_{0}=0.3$, $\Lambda_{0}=0.7$ (bold lines).
The 
horizontal axis has been labeled both in units of velocity ($w_{s}$) and 
comoving space separation $s$, calculated for the fiducial 
model.
 The observed correlation function is given by the monopole term in the 
Legendre polynomial expansion, $C_{0}(\beta,w_{s})$. In order to match the  
value of the $C_{0}$ coefficient to unity at the observed correlation length of
$2.1 h^{-1} \mpc$ (comoving) for $\Omega_{0}=1$ (Giavalisco \etal 1998), 
which corresponds to a correlation 
velocity of $450 \kms$, the bias required is $b=4$. 
For the 
$\Omega_{0}=0.3$, $\Lambda_{0}=0.7$ model, the bias required to match the 
computed $C_{0}$ coefficient to $1$ at the observed correlation velocity of
$450 \kms$ is $2.3$. On account of the large bias in these two models, we
can rely on the linear theory that we have used for peculiar velocity 
distortions of the correlation function we have shown in \S 
3 for $C_{0} 
\lesssim 1$, or $w_{s} \gtrsim 400 \kms$. 

Our goal is to measure the multipoles $C_{l}(\beta,w_{s})$ of the 
correlation function and use it to constrain the geometric distortion 
parameter, $\alpha$. From Figure 4 we see that on scales of
approximately $10^{3} \kms$, the $C_{2}$ coefficient 
is a $10 \%$ perturbation on the monopole term, whereas the octapolar 
term $C_{4}$ is a smaller contribution at $\sim 3 \%$ for the
$\Lambda_{0}=0.7$ model. 
Once we fix the bias, the $C_{0}$ 
coefficients for the two models shown are similar to each other, except on
very large scales.
The quadrupolar coefficients for the two models on the other hand are 
very different. The $C_{2}$ coefficient for the $\Lambda$ model, affected 
by geometric distortion, is larger 
by a factor of 2 compared to the $\Omega_{0}=1$ model and comes to within a 
factor of 2 of the monopole term on scales $\sim 10^{4} \kms$. As we mentioned
in \S 2, the $C_{2}$ coefficient is less than zero which implies a 
squashing of the 
contours of constant $\rxi$ along the line of sight.
Thus we see that for our choice of the fiducial model, the primary effect
of geometric distortion caused by a model with a positive cosmological 
constant is to cause a further squashing of the contours of $\rxi$. 
We mention here that the $C_{4}$ coefficient is even more sensitive to 
geometric distortion than the $C_{2}$ coefficient. Its value is 
approximately $10$ times larger for the $\Lambda_{0}=0.7$ model as compared 
to the fiducial $\Lambda=0$ model. A measurement of the $C_{4}$ coefficient 
will give us additional information with which to test the bias model that 
we have used. It will be interesting  to compare the value of the 
linear bias parameter parameter derived from a simultaneous measurement 
of the cosmological constant and the $\beta$ parameter using both the $C_{4}$
and $C_{2}$ coefficients, to the value obtained by comparison of the 
galaxy distribution to the matter power spectrum at redshift zero.  

We now show that the difference in angular distortion of the correlation
function in the two models is 
primarily due to the change in the distortion parameter and is not
strongly dependent on the choice of the power spectrum.
In Figure 5, we plot the coefficients $C_{l}$, for the fixed cosmological
model $\Omega_{0}=0.3,\Lambda_{0}=0.7$, but two different correlation functions.
The bold lines correspond to the power spectrum of the $\Lambda$ model with
these same parameters. 
The lighter lines are for the same cosmological model, but with the
power spectrum of an $\Omega_{0}=1$
CDM model as a function of $k/H(z)$. 
We see from this figure that
at velocity separations $ \gtrsim 10^4 \kms$, the differences in the 
power spectra dominate the differences in the $C_{l}$ coefficients. But on 
smaller scales the geometric distortion effect is the
most important effect. In particular the $C_{2}$ and $C_{4}$ coefficients 
are similar once the monopoles for both models are normalized to unity at
the observed correlation length. 
Thus the ratios of the coefficients 
$C_{2}/C_{0}$ and $C_{4}/C_{0}$ are only weakly dependent on the shape 
of the power spectrum. This shows that it should be possible to measure
the 
geometric distortion parameter even if the power spectrum is not known 
accurately from independent methods.

\section{Error estimates}
In this section we compute the accuracy in the measurement of the
multipoles of the redshift space correlation function from a typical 
survey volume and test the feasibility of the method described above.
Currently, the typical observed fields have a size $\sim 12 \arcmin$ on
each side. 
The redshift range of each field extends from $z=2.6$ to $z=3.4$  with a
surface density of approximately $1.25$ Lyman-break objects per square 
arc minute within this redshift range (Adelberger \etal 1998).
In our fiducial 
model ($\Omega_{0}=1$), this corresponds to a width of $ 2 \times  10^{3} \kms$ and a depth of $ 6 \times 10^{4} \kms$. 
We consider for the purpose of  
error estimation, a wide field of view of $\sim 3^{\circ}$. We shall later 
discuss the scaling of the errors with the angular size of the field of view.

 Any detailed calculation of the 
errors in a survey will depend upon the  precise geometry of the survey
volume and the selection effects involved in the survey. Here, we consider 
two of the sources of error, shot noise and cosmic variance.
Shot noise is caused by the discrete nature of the 
galaxies from which we 
measure the correlation function. Cosmic variance arises due to the finite volume
we use to estimate a statistical quantity. 
We calculate these errors for  
a single cylindrical survey 
volume with a radius of $ 1.5^{\circ}$ ($75 h^{-1} \mpc$ for $\Omega_{0}=1$
 model), and depth extending from $z=2.6$ to $z=3.4$ ($300 h^{-1} \mpc$ for 
$\Omega_{0}=1$ model).  
At the current estimate of surface density of Lyman-break galaxies of $1.25$ per 
square arcmin (Adelberger \etal 1998), approximately $30000$ galaxies would be 
included in our survey volume. 

\subsection{Shot noise}

In order to estimate the redshift space correlation function, we bin pairs of
galaxies with respect to their separation velocity $w_{s}$ (computed in the 
fiducial model)  in widths of $\Delta w_{s}$. The redshift correlation 
function is then estimated (denoted by subscript E) as, 
\begin{equation}
\tilde{\xi}_{E}({\bf w}_{s})=\frac{N_{p}(w_{s},\mu_{s})}{\overline{N}_{p}(w_{s},\mu_{s})}-1,
\label{eqnrxi}
\end{equation}
where $N_{p}(w_{s},\mu_{s})$ are the number of pairs with 
separations between  $w_{s}$ and $\Delta w_{s}$ with the separation vector 
making  an angle ${\rm cos}^{-1}(\mu_{s})$ with the line-of-sight and 
$\overline{N}_{p}(w_{s})$ is the ensemble average of a random distribution of 
the same quantity. There are various different estimators for the correlation 
function discussed in the  literature that minimize the error of the
estimator due to the unknown true average density of the galaxies at the 
redshift of the survey (for a discussion see Hamilton 1993). 
Our shot noise will be dominated by the small number of pairs of galaxies we 
have in each of our bins.
Since we are currently only interested in an estimate of this error,
we have adopted the simpler estimator for the correlation function. To analyze
the data from a survey one should use a more sophisticated estimator to minimize
its variance. 

Using equation (\ref{eqnrxi}) we obtain the  estimate of the $C_{l}$ coefficients as given below for $l \neq 0$  
\begin{equation}
C_{l,E}(w_{s})=\frac{2l+1}{2} \frac{1}{\overline{N}_{p}(w_{s})} \cdot  \sum_{i=1}^{N_{p}} P_{l}(\mu_{si}) \,,
\label{eqnClp}
\end{equation}
where $N_{p}$ is the number of pairs with separations between $w_{s}- \Delta w_{s}$ 
and $w_{s}+ \Delta w_{s}$, and  $\overline{N}_{p}(w_{s},\mu_{s})$ is the average 
number of pairs for a random distribution of galaxies in the same bin .
The summation is performed over the pairs of galaxies (denoted by subscript i) 
in the bin centered at $w_{s}$.
In order to calculate the statistical average of the estimator, we have to
perform two integrals. First, for a given number of pairs separated by 
$w_{s}$, we average over their possible orientations.  The probability that  a given 
pair of galaxies with separation $w_{s}$ is oriented along $\mu_{s}$ is given by 
$\psi(w_{s},\mu_{s})$, where
\begin{equation}
\psi(w_{s},\mu_{s})  d\mu_{s}=\frac{1+ \tilde{\xi}(w_{s},\mu_{s})}{1+C_{0}(w_{s})}  d\mu_{s}.
\end{equation}
The $1+C_{0}(w_{s})$ factor in the  denominator comes from normalizing 
$1+ \tilde{ \xi}(w_{s},\mu_{s})$ over $\mu_{s}$.  Here we have assumed that a 
given 
pair of galaxies can  have any orientation with respect to the line of sight. 
This is 
clearly not true, for example for a pair of galaxies close to the edge of the 
survey 
volume. In order the circumvent this difficulty we consider a smaller volume 
within 
the total survey volume which we call the ``reduced volume'', hereafter 
denoted as 
$V_{R}$ such that the edges of $V_{R}$ are a distance $w_{s}$ away from 
the 
edges of the total survey volume.  We only consider pairs of galaxies such 
that at 
least one 
of the galaxies is within $V_{R}$ .  For a random distribution of galaxies, a 
pair chosen in this way is not biased to be aligned along a particular
direction. We can  
see that the largest separation at which we can measure  the coefficients $C_{l}$ 
is  the radius of the survey for which $V_{R}$ goes to zero. 

Secondly, we have to average  over the  distribution
of the number of pairs of galaxies in each bin.  
Calculating the averages (denoted by brackets) yields
\begin{equation}
<C_{l,E}(w_{s})>= \frac{2l+1}{2} \frac{<N_{p}(w_{s})>}{\overline{N}_{p}(w_{s})} \int  \psi(w_{s},\mu_{s}) P_{l}(\mu_{s}) d \mu_{s} \,.
\end{equation}
This gives us that $<C_{l,E}(w_{s})> = C_{l}(w_{s})$. This result can also 
be shown to hold for the monopole term.

In a similar way to the calculation of the statistical average of the $C_{l}$
coefficients, we can calculate the mean square variation of $C_{l}$ 
coefficients. Using equation (\ref{eqnClp}) we have,
\begin{equation}
C_{l,E}^{2}(w_{s})=\left( \frac{2l+1}{2} \right)^{2}\left( \frac{1}{\overline{N}_{p}} \right)^{2} \left( \sum_{i=1}^{N_{p}} P_{l}(\mu_{si})^{2} + 2 \sum_{i<j}^{N_{p}} P_{l}(\mu_{si})P_{l}(\mu_{sj}) \right).
\label{Clsqpoiss}
\end{equation}

The statistical average of the above equation gives the mean square variance 
of the $C_{l}$ coefficients, $<C_{l,E}^{2} - <C_{l,E}^{2}>>$, denoted by $\sigma_{l}^{2}$,
as,
\begin{equation}
\sigma_{l}^{2}(w_{s})=\left( \frac{2l+1}{2} \right) \frac{(1+C_{0}(w_{s})) }{\overline{N}_{p}(w_{s})}.
\end{equation}

We mention here that in deriving the above equation we have assumed a Poisson
distribution for the number of pairs in each bin. This assumption is not strictly
valid since every pair separation is not independent. Hence, one may expect
some underestimation in the Poisson errors we have calculated but this should
be small since the second term in equation (\ref{Clsqpoiss}) is proportional
to $C_{l}^{2}$. 
Figure 6 shows the expected $1 \sigma$ error for the $C_{l}$ coefficients 
due to shot noise. Each successive  bin is centered at $w_{s}$ with value $1.5$ times
that of the previous bin and hence, each bin has  width $2/5 w_{s}$. The average 
number of pairs $\overline{N}_{p}(w_{s})$ in the bin centered at 
$w_{s}$ and width $2 \times \Delta w_{s}$ 
for a random distribution of galaxies within the survey volume is given by 
\begin{equation}
\overline{N}_{p}(w_{s})=\frac{\overline{n}_{g}^{2}}{2} \times {\rm V_{R}} \times 4 \pi w_{s}^{2} 2 \Delta w_{s} \,,
\end{equation}
where $\overline{n}_{g}$ is the average density of galaxies within the survey 
volume. This is an underestimate of the number of pairs in the bin since it counts 
only half of the pairs of galaxies of which one of the galaxies is outside 
$V_{R}$.
At larger separations this underestimation is maximum since, in this case, 
a larger fraction of all the pairs in the bin have one of the galaxies outside of 
 $V_{R}$. Thus our shot noise is an overestimate by a factor $\leq
\sqrt{2}$. 

We can see from Figure $6$ that with shot noise alone, the $C_{l}$ coefficients 
are 
best measured in the velocity range $10^{2} \lesssim w_{s} \lesssim10^{4} \kms$ 
for a survey of the size and geometry that we have assumed. The 
errors on the multipoles scale as $(2l+1)^{\frac{1}{2}}$, and so they are 
smaller for
the $C_{0}$ coefficient and higher for the $C_{4}$ coefficient as compared to the 
quadrupole. For scales close to the radius of the survey the shot noise 
error increases 
rapidly since $V_{R}$ is now very small.  
For scales 
$\sim 10^{3} \kms$, with shot noise alone, we can measure the $C_{2}$ coefficient 
to a few percent accuracy, both for the fiducial model as well as the $\Lambda$ 
model and hence distinguish a large cosmological 
constant as in our model with $\Lambda_{0} = 0.7$ to high statistical significance.
The shot noise error on the $C_{4}$ coefficient is small for the 
$\Lambda_{0}=0.7$ model we have shown but larger for models with smaller 
cosmological constants. Considering shot noise alone, on scales of $\sim\, 10^{3} \kms$, the $C_{4}$ coefficient can be measured if it is present at the 
level of a few percent of the monopole, which in turn would indicate a 
large energy density in the form of a cosmological constant or some form
of quintessence.
We also note here that the number of pairs of galaxies at a fixed separation is 
proportional to $V_{R}$. Hence for separations small compared to the 
radius of the survey,  the shot noise error scales as the inverse of the angular 
size of the survey.

\subsection{Cosmic variance}
The cosmic variance of a survey volume results from the sparse sampling 
of the universe made by the small survey volume. It occurs even if the 
overdensity at each point within the survey volume is accurately known,
and is independent of the number of observed galaxies. We estimate the 
cosmic variance in this section using linear theory.

A finite volume estimate (denoted by subscript E) of the correlation function is given 
by,
\begin{equation}
\tilde{\xi}_{E}(\vec{w_{s}},\hat{n}) = \frac{1}{{\rm V_{R}}} \int_{{\rm V_{R}}} d^{3}x \, \delta (\vec{x},\hat{n}) \delta (\vec{x} + \vec{w_{s}},\hat{n}) \, .
\label{eqnxiE}
\end{equation}
In the above equation, $\vec{x}$ is constrained to be within $V_{R}$ such that its 
boundary are a distance $w_{s}$ away from that of the full survey volume and 
$\vec{x} + \vec{w_{s}}$ is within the full survey volume. 
For every point $\vec{x}$ within 
$V_{R}$, the overdensities at $\vec{x}$
and $\vec{x}+\vec{w_{s}}$ are 
accurately known. The ensemble average of the estimator gives,
\begin{eqnarray}
< \tilde{\xi}_{E}(\vec{s})> = \frac{1}{{\rm V_{R}}} \int_{{\rm V_{R}}} d^{3}x \, \tilde{\xi}(\vec{s}) \, ,  \nonumber \\
=\tilde{\xi}(\vec{s}) \, ,
\end{eqnarray}
where as previously, quantities without the subscript $E$ stand for their 
true values. 
Similarly,
\begin{equation}
<C_{l,E}(w_{s})>=C_{l}(w_{s}).
\end{equation}
The variance in $C_{l,E}(w_{s})$ can be computed using,
\begin{equation}
<C_{l,E}^{2}(w_{s})>=\left( \frac{2l+1}{2} \right)^{2} \int d\mu_{s1} \int d\mu_{s2} < \tilde{\xi}_{E}(s,\mu_{s1}) \tilde{\xi}_{E}(s,\mu_{s2}) > P_{l}(\mu_{s1}) P_{l}(\mu_{s2}) \, ,
\label{eqnCl}
\end{equation}
where,
\begin{equation}
< \tilde{\xi}_{E}(w_{s},\mu_{s1}) \tilde{\xi}_{E}(w_{s},\mu_{s2}) > =\frac{1}{{(\rm V_{R}})^{2}} \int d^{3}x_{1} \int d^{3}x_{2} < \delta (\vec{x_{1}}) \delta (\vec{x_{1}}+\vec{w_{s1}}) \delta (\vec{x_{2}})  \delta (\vec{x_{2}}+\vec{w_{s2}})> \,,
\end{equation}
where $|\vec{w}_{s1}|=|\vec{w}_{s2}|$ and $\hat{w}_{s1} \cdot \hat{n}= \mu_{s1}$, $\hat{w}_{s2} \cdot \hat{n}= \mu_{s2}$. 

In order to simplify the above expression, we approximate the overdensities
to be in the linear regime. The linear overdensities are 
Gaussian distributed and the four point expression in the above equation can be 
expressed in terms of two point correlation functions :
\begin{eqnarray}
<\delta (\vec{x_{1}}) \delta (\vec{x}_{1}+\vec{w}_{w1}) \delta (\vec{x}_{2})  \delta (\vec{x}_{2}+ \vec{w}_{s2}) > = 
\tilde{\xi} ( \vec{w}_{s1}) \tilde{\xi} ( \vec{w}_{s2}) \nonumber \\
+ \tilde{\xi} ( \vec{x}_{1}- \vec{x}_{2}) \tilde{\xi} ( \vec{x}_{1}- \vec{x}_{2}+ \vec{w}_{s1}- \vec{w}_{s2})  \nonumber \\
+ \tilde{\xi} ( \vec{x}_{1}- \vec{x}_{2}+ \vec{w}_{s1}) \tilde{\xi} ( \vec{x}_{1}- \vec{x}_{2}- \vec{w}_{s2}).
\label{eqn4pt}
\end{eqnarray}
Thus the last two terms in equation (\ref{eqn4pt}) contribute to the root mean 
square variance in $C_{l,E}(w_{s})$. Since $\tilde{\xi}_{E}(w_{s},\mu_{s})$ is 
independent of the azimuthal angle in equation (\ref{eqnCl}), we  can also 
integrate  over this angle.  Therefore we can express the root mean square 
variance as,
\begin{eqnarray}
<\sigma_{C_{l,E}}^{2}(w_{s})>=\left(\frac{2l+1}{2 {\rm V_{R}}} \right)^{2} \int d^{3}x_{1} \int d^{3}x_{2} \frac{d \Omega_{1}}{2 \pi} \frac{ d \Omega_{2}}{2 \pi} P_{l}(\mu_{s1}) P_{l}(\mu_{s2}) \nonumber \\
\left\{ \tilde{\xi} ( \vec{x}_{1}- \vec{x}_{2}+ \vec{w}_{s1}) \tilde{\xi} ( \vec{x}_{1}- \vec{x}_{2}- \vec{w}_{s2}) + 
\tilde{\xi} ( \vec{x_{1}}- \vec{x_{2}}) \tilde{\xi} ( \vec{x}_{1}- \vec{x}_{2}+ \vec{w}_{s1}- \vec{w}_{s2}) \right\}. 
\label{eqnsigma}
\end{eqnarray}

The method we employed in the calculation of the above integrals is detailed in 
the Appendix. 
Figure 7 displays the expected cosmic variance errors for our survey 
volume for the multipole coefficients. For all three coefficients, the 
cosmic variance  dominates the error on large scales, while the shot noise 
contribution is larger on smaller scales.
The error is smallest in the region  
$w_{s} \sim 3000 \kms$, so this is the best scale at which to measure the 
quadrupole and octapole coefficients and hence estimate the geometric 
distortion factor. 
 
As mentioned before, the cosmic variance error that we have calculated assumes 
linear theory and hence we have underestimated the contribution to the errors from 
fluctuations 
and non-linear velocity effects on small scales. 
As mentioned in \S 3 if we adopt 
a local but stochastic model for the distribution of 
galaxy number density as a function of the underlying mass density, then the bias 
will still be scale independent on large scales, but there will be a larger variance in 
the measured 
correlation function. In the absence of a well motivated stochastic biasing model, we have not
estimated the variance in the $C_{l}$ coefficients arising from such a model of the bias. 
Depending on the true nature of bias, we may be underestimating the variance of the 
measured correlation function. 
A more precise estimate of the error can only be given by the direct analysis of 
numerical simulations, a project we plan to return to in a later paper. 

We note here that we have not assumed that the mean overdensity within the 
survey volume is zero. The fluctuation in the mean overdensity is the 
primary source of the cosmic variance error for the monopole component of the 
correlation 
function. This fluctuation of course does not affect the higher multipoles of 
the
correlation function and hence on small scales the error on the higher 
multipoles
is smaller than on the monopole coefficient. 
With the combined shot noise and cosmic variance errors the $C_{2}$ 
coefficient can be measured to a few percent accuracy both for the 
fiducial $\Omega_{0}=1$ model as well as the $\Lambda$ models which have 
a larger quadrupole coefficient compared to the fiducial model.   
Thus, with our estimate of 
the errors, we can distinguish a geometric distortion factor of about $15 \%$ 
corresponding to a $\Lambda$ model with $\Lambda_{0}=0.7$ to high statistical 
significance.
From Figures 6 and 7 we also see that for our 
survey volume, the $C_{4}$ coefficient can be measured to $\sim 20 \%$ 
accuracy for the $\Lambda_{0}=0.7$ Model. The errors are larger for models 
with smaller cosmological constants. Therefore, a measurement of
the 
octapolar coefficient is possible if it is present at the level of a 
few percent of the monopole on scales of $\sim 3\times 10^{3} \kms$ 
as in case of a large cosmological constant. Since the error on $C_{4}$ is 
large, a simultaneous measurement of both the $\beta$ parameter as well as 
the cosmological constant from the anisotropy of the redshift space 
correlation function alone is difficult. This has been indicated earlier by 
Ballinger \etal (1996). If the the octapolar coefficient
can be measured, and the bias parameter constrained, it will be interesting 
to  compare its value to the one obtained by comparison of galaxy clustering
to the assumed underlying matter distribution. 
But we emphasize that when we assume that the amplitude of the 
matter power spectrum is known, and only 
one parameter needs to be measured from the redshift space correlation 
function, then the quadrupolar geometric distortion effect of the 
cosmological constant can be measured to high accuracy. 

For scales much smaller than the 
radius of the survey, the cosmic variance error scales as the inverse square root of the 
volume of the survey and hence as the inverse of the 
angular size of the survey. Thus both shot noise and cosmic variance have similar 
dependence on the angular size of the survey on small length scales. A large 
cosmological constant may be distinguished with high statistical significance for 
smaller angular size surveys depending upon other sources of error. Considering
only the shot noise and the cosmic variance that we have estimated, 
for a survey of angular size  $1^{\circ}$, 
with a factor of three 
increase in the errors, we can still measure the quadrupolar coefficient
affected by a geometric distortion parameter of $15 \%$ with an accuracy 
of approximately $10 \%$ on a scale of $3 \times 10^{3}\kms$. The error is 
larger for smaller distortion factors. Since a variation of the
cosmological constant from zero to $0.7$ changes the quadrupolar
coefficient by a factor of 2, we can use a linear relation between the two
to make an approximate estimate of the accuracy with which the value 
of the cosmological constant can be measured. This gives us that 
a large cosmological constant, for which the error in the difference of the
$C_{2}$ coefficient with respect to its  value in the fiducial model is small,  
can be constrained with an error bar of
approximately $20 \%$ with a $1^{\circ}$ field of view.  Since in 
fact this linear relation is incorrect and the 
geometric distortion parameter is more sensitive to a variation in the  
cosmological constant when it is large (Ballinger \etal 1996), the error
we have quoted will be somewhat smaller for large $\Lambda_{0}$ 
($\gtrsim 0.5$). 
For a field of view of this size, 
the $C_{4}$ coefficient can also be measured, although with a large error 
of $\sim 60 \%$, if it is present at the level of a few percent of the
monopole as in the case of geometric distortion with respect to the 
fiducial $\Omega_{0}=1$ model by a cosmological constant
$\Lambda_{0}=0.7$. 

For a smaller field of view, the monopole coefficients have to be measured 
on scales smaller than $3000 \kms$ where shot noise is the dominant source 
of error. For example for a field size of $1/2^{\circ}$, the quadrupole 
coefficient corresponding to a $15 \% $ geometric distortion parameter 
can still be measured to an accuracy of approximately $50 \% $ on a scale of 
$10^{3} \kms$. Hence it can be distinguished from the
fiducial $\Omega_{0}=1$ model at the $2 \sigma$ level. For smaller scales 
the error is larger while to measure the distortion parameter at larger 
scales a larger field size is required.  Thus a field at least
$1/2^{\circ}$ in diameter, corresponding to an area approximately 
four times the currently used field size, is required to distinguish a 
$\Lambda_{0}=0.7$ model from our fiducial $\Omega_{0}=1$ model.

\section{Discussion and Conclusions}
In this paper we have investigated the feasibility of using the high redshift 
population of Lyman-break galaxies to measure the geometric distortion effect 
and hence constrain cosmological parameters. The method is particularly sensitive 
to components of energy density with negative pressure and in particular 
to the cosmological constant. 
The principal advantage of using this population of
galaxies 
is their high bias with respect to the underlying matter distribution. This 
tends to suppress the peculiar velocity effects and makes it easier to measure the 
geometric distortion effect. As pointed out by Ballinger \etal (1996), a 
simultaneous 
measurement of the bias and the cosmological constant using the 
redshift space distortion alone is difficult except in case of a large cosmological
constant. 
In this paper we assumed that the matter power spectrum at redshift 
$3$ is related by the linear growth factor to the matter power spectrum at 
redshift zero which is constrained by observations of cluster abundances.
We fixed the bias of the Lyman-break galaxies by comparing their clustering 
to the assumed matter power spectrum at redshift $3$. Then we only need to 
measure one parameter, the geometric distortion parameter, from the 
anisotropy of the 
correlation function.  
This permits us to use the lowest order quadrupolar
distortion of the redshift space power spectrum to constrain the geometric 
distortion parameter to high accuracy. In cases of a large energy density in
a cosmological constant or quintessence, the octapolar 
coefficient may also be measured. An interesting test would then be to 
compare the
value of the bias parameter derived from the additional information provided 
by the octapolar term to that determined by comparing the galaxy clustering
to the matter power spectrum.

We estimated that in order to distinguish a flat 
model with 
$\Lambda_{0}=0.7$ from the Einstein-de Sitter case,  at least a $1/2^{\circ}$ 
sized circular field of view
is required. Currently
the observation fields have sizes of approximately $10^{'}$, which are too small 
for measurements of geometric distortion, both due to shot noise and
cosmic variance. It is preferable to measure the distortion effect on 
large 
scales where the effects of peculiar velocities can be analytically computed
using linear theory. 
For this  reason, it is better to use a single large 
field of view than to combine data from several small fields of view which 
provide data only on smaller scales. 
For a more accurate measurement of the distortion parameter larger field 
sizes are required.
We estimated that for a field size of $3^{\circ}$, the best scale at which to 
measure the ratio of the quadrupole 
coefficient to the monopole is approximately $3000 \kms$, or $15 h^{-1} \mpc$ 
in the $\Omega_{0}=1$ model and somewhat smaller for a smaller field.
Since the difference in the quadrupolar coefficients for the 
flat $\Lambda_{0}=0$ and $\Lambda_{0}=0.7$ models can be measured to 
$\sim 20 \%$ accuracy with a circular field of diameter $1^{\circ}$, we made a 
rough estimate that a large cosmological constant $\gtrsim 0.5$ can be measured
with this precision.

Our cosmic variance was estimated using the linear 
correlation function and we have underestimated the error due to fluctuations and non linear 
velocity effects on small scales.
We have also used a very simple local non-stochastic scale 
independent model for the bias. Stochastic bias will lead to variance in the measured 
correlation function which we have not accounted for.  A full calculation
of the errors including non linear effects will require analysis of numerical 
simulations, which we will discuss
in a future paper.  

\acknowledgements 
I wish to acknowledge Jordi Miralda-Escud\'e, my thesis advisor, who gave me 
the original motivation
for this work and for the numerous insightful comments and 
discussions I have had  with him. I also wish to thank Patrick MacDonald, Brian
Mason and David Moroz for their comments on the paper. I would also like to acknowledge
the anonymous referee for his comments and suggestions that have improved the content and
presentation of the paper.


\appendix
\setcounter{secnumdepth}{0}
\section{ Appendix}
\subsection{ Calculation of integrals for the cosmic variance}
\setcounter{section}{1}

We perform the integrals required in equation (\ref{eqnsigma}) for the case 
of a cylindrical volume of radius R and length $L_{z}$. 
Since equation (\ref{eqn4pt}) 
depends only on the difference vector $\vec{x_{1}} -\vec{x_{2}}$, 
the six dimensional integral over $\vec{x_{1}}$ and $\vec{x_{2}}$ can be reduced to a two dimensional integral. We define sum and 
difference vectors $\vec{x}_{+}$ and $ \vec{x}_{-}$ respectively as 
\begin{eqnarray}
\vec{x}_{+} = \vec{x}_{1} + \vec{x}_{2} \nonumber \\
\vec{x}_{-} = \vec{x}_{1} - \vec{x}_{2}, 
\end{eqnarray}

Denoting our integrand as $f( \vec{x}_{-})$ we have the following result.
\begin{equation}
\int \int d^{3}x_{1} d^{3}x_{2} f( \vec{x}_{-}) = \frac{1}{8} \int d^{3}x_{-} \, 
V_{+}( \vec{x}_{-}) \, f( \vec{x_{-}}),
\end{equation}
where $V_{+}(\vec{x}_{-})$ is the volume occupied by the sum vector $\vec{x}_{+}$ for 
a fixed difference vector $\vec{x}_{-}$. Denoting the components of the $\vec{x}_{-}$
in cylindrical coordinates as $\rho_{-}$ and $z_{-}$,  $V_{+}(\vec{x}_{-})$ is 
given as
\begin{equation}
V_{+}(\vec{x}_{-})=\{ 2 {\rm R}^{2} cos^{-1}( \frac{\rho_{-}}{2{\rm R}}) - \rho_{-} \left({\rm R}^{2} - \frac{ \rho_{-}^{2}}{4} \right)^{\frac{1}{2}} \} \cdot 2 ( {\rm L_{z}}-|z_{-}|).
\end{equation}

Let us first consider the contribution of the first term in equation (\ref{eqnsigma}), 
 $\tilde{\xi}( \vec{x}_{-}+ \vec{w}) \tilde{\xi} ( \vec{x}_{-}- \vec{w}_{2})$, and denote
it by $<\sigma^{2}_{C_{l,E}}(w)>_{I}$:
\begin{equation}
 <\sigma^{2}_{C_{l,E}}(w)>_{I}= \left( \frac{2l+1}{2} \right)^{2} \int d^{3}x_{1} \int d^{3}x_{2} \left(I_{l}(\vec{x}_{-},w) \right)^{2} \, ,
\end{equation}
where,
\begin{equation}
I_{l}(\vec{x}_{-},w)= \int \frac{d \Omega_{1}}{2 \pi} P_{l}( \mu_{1}) \tilde{\xi}(\vec{x}_{-} + \vec{w}_{1}) \, .
\end{equation}
We now calculate $I_{l}(\vec{x}_{-},w)$. The Fourier transform of $\tilde{\xi}$ is 
\begin{equation}
\tilde{\xi}(\vec{x}_{-}+ \vec{w}_{1})=\frac{1}{(2 \pi)^{3}} \int d^{3}k \tilde{P}(k,\hat{k} \cdot \hat{n}) e^{i \vec{k} \cdot (\vec{x}_{-}+ \vec{w_{1}})},
\end{equation}
where $\tilde{P}(k,\hat{k} \cdot \hat{n})$ is the redshift space power spectrum
given by, 
\begin{equation}
\tilde{P}(k,\hat{k} \cdot \hat{n})=P(k) \sum_{l=0,2,4} (-1)^{l}  A_{l}(\beta) P_{l}(\hat{k} \cdot \hat{n}) \, .
\label{eqnPk} 
\end{equation}
The coefficients $A_{l}(\beta)$ are defined in equation (\ref{eqnAl}).
For convenience of computation, we take the line-of-sight vector $\hat{n}$ 
to lie along the z axis. 
We first perform the integrals over the angles $\Omega_{1}$ and $\Omega_{2}$ in equation 
(\ref{eqnsigma}). Using,
\begin{equation}
e^{i \vec{k} \cdot \vec{w}_{1}}=4 \pi \sum_{L,M} i^{-L} j_{L}(kw) Y_{LM}( 
\Omega_{k}) Y_{LM}^{*}( \Omega_{w_{1}}),
\end{equation}
we have,
\begin{equation}
\int \frac{d \Omega_{1}}{2 \pi} e^{i \vec{k} \cdot \vec{w_{1}}} P_{l}( \mu_{1})= 2 i^{-l} j_{l}(kw) P_{l}(\mu_{k}),
\end{equation}
where $\mu_{k}=cos(\hat{k} \cdot \hat{n})$. Thus,
\begin{equation}
I_{l}(\vec{x}_{-},w)=
\frac{i^{-l}}{4 \pi^{3}} \int d^{3}k \tilde{P}(k,\hat{k} \cdot \hat{n}) e^{i \vec{k} \cdot \vec{x}_{-}} j_{l}(kw) P_{l}(\mu_{k}).
\end{equation}
Substituting equation (\ref{eqnPk}) in  above equation And integrating over $d^{3}k$ ,
\begin{equation}
I_{l}(\vec{x}_{-},w)=
\frac{i^{-l}}{2 \pi^{2}} \sum_{l^{'}=0}^{8} i^{-l^{'}} (2l^{'}+1) D1(l,l^{'}) 
\chi(l,w,l^{'},x_{-})P_{l^{'}}(cos \theta_{x_{-}}),
\end{equation}
where,
\begin{eqnarray}
D1(l,l')=\sum_{l^{''}=0,2,4} A_{l^{''}}( \beta) B1(l,l^{'},l^{''}), \nonumber \\
B1(l,l^{'},l^{''})= \int d \mu_{k} \, P_{l} \left( \mu_{k} \right) \,
 P_{l^{'}} \left( \mu_{k} \right) \, P_{l^{''}} \left( \mu_{k} \right), \\
\end{eqnarray}
and
\begin{equation}
\chi(l,w,l^{'},x_{-})= \int dk \, k^{2} \, j_{l}(kw) j_{l^{'}}(k x_{-}) P(k).
\end{equation}

The contribution of the second term in equation. (\ref{eqnsigma}) to $<\sigma_{C_{l,E}}^{2}(w)>$ can be computed in a similar fashion.

\newpage
\begin{figure}
\centerline{
\hbox{
\epsfxsize=4.4truein
\epsfbox[55 32 525 706]{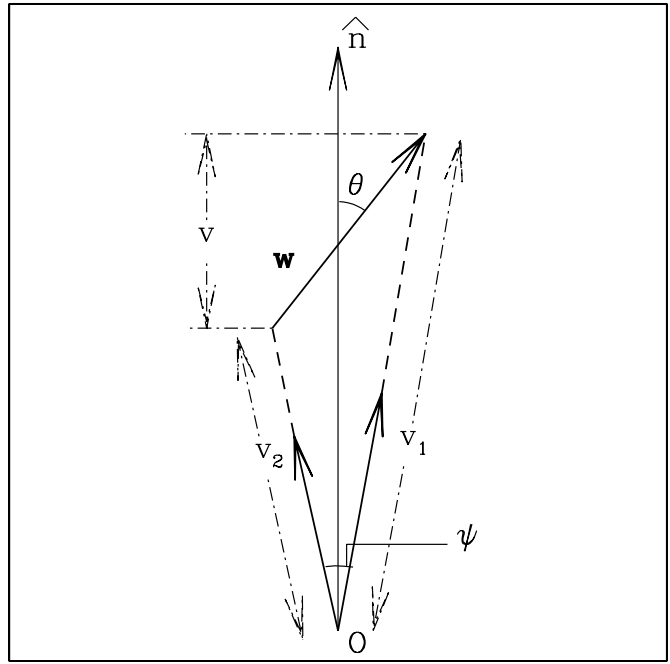}
}
}
\vskip -40pt
\caption{
v1 and v2 are observed velocities of two objects along the line-of-sight,
with a small angular separation $\psi$. Their separation is velocity space
is {\bf w} which makes an angle $\theta$ with the line-of-sight. The velocity
separation along the line of sight is $v$ and perpendicular to the 
line-of-sight is $H(z)\,D(z)\,\psi$.
}
\end{figure}
\vfill\eject
\newpage

\begin{figure}
\centerline{
\hbox{
\epsfxsize=4.4truein
\epsfbox[55 32 525 706]{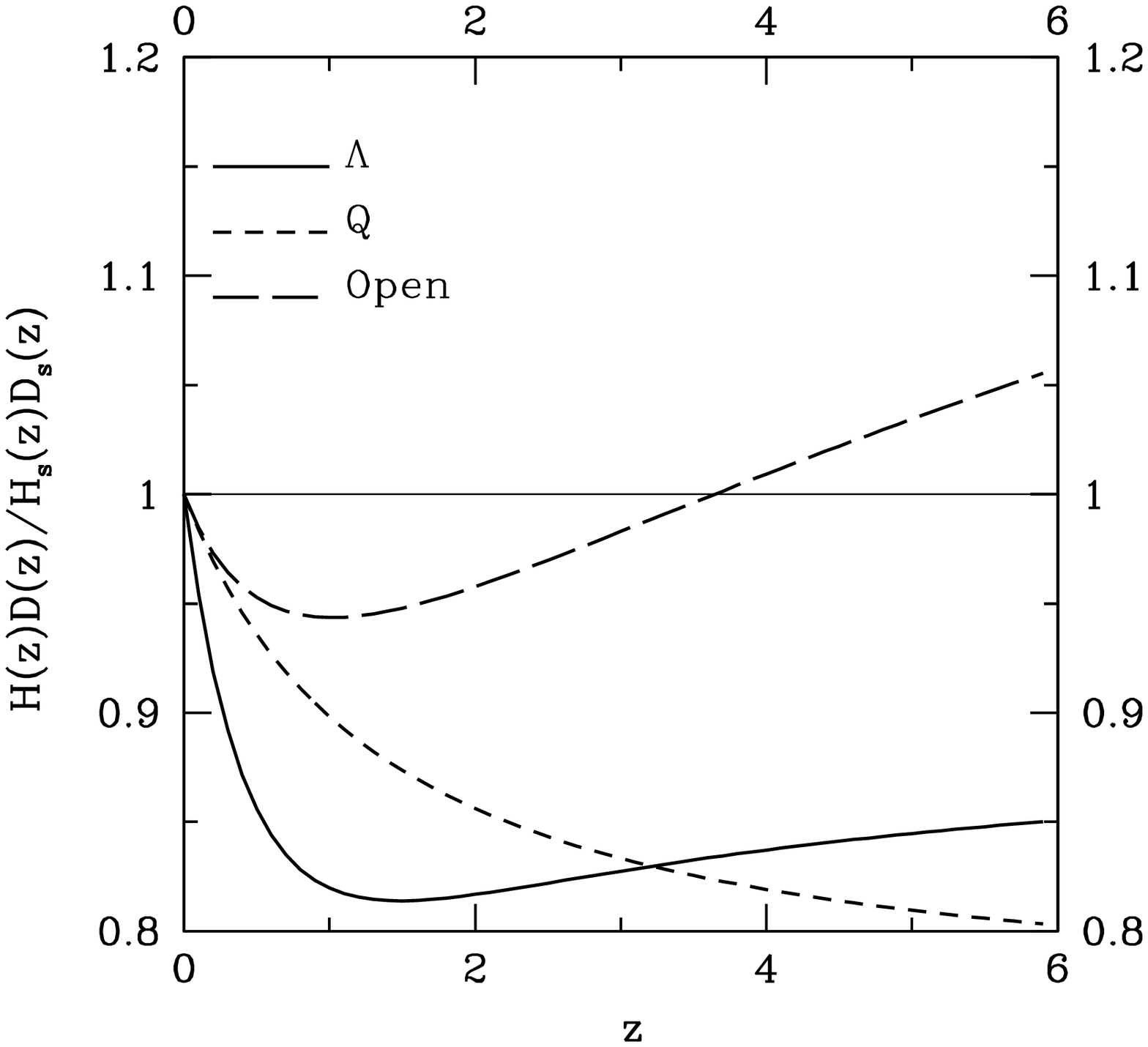}
}
}
\vskip -40pt
\caption{ Ratio of $H(z)\,D(z)$ in a Model to its value in the fiducial
$\Omega_{0}=1$ Model. $\Omega_{0}=0.3$ for all three Models.
}
\end{figure}
\vfill\eject
\newpage

\begin{figure}
\centerline{
\hbox{
\epsfxsize=4.4truein
\epsfbox[55 32 525 706]{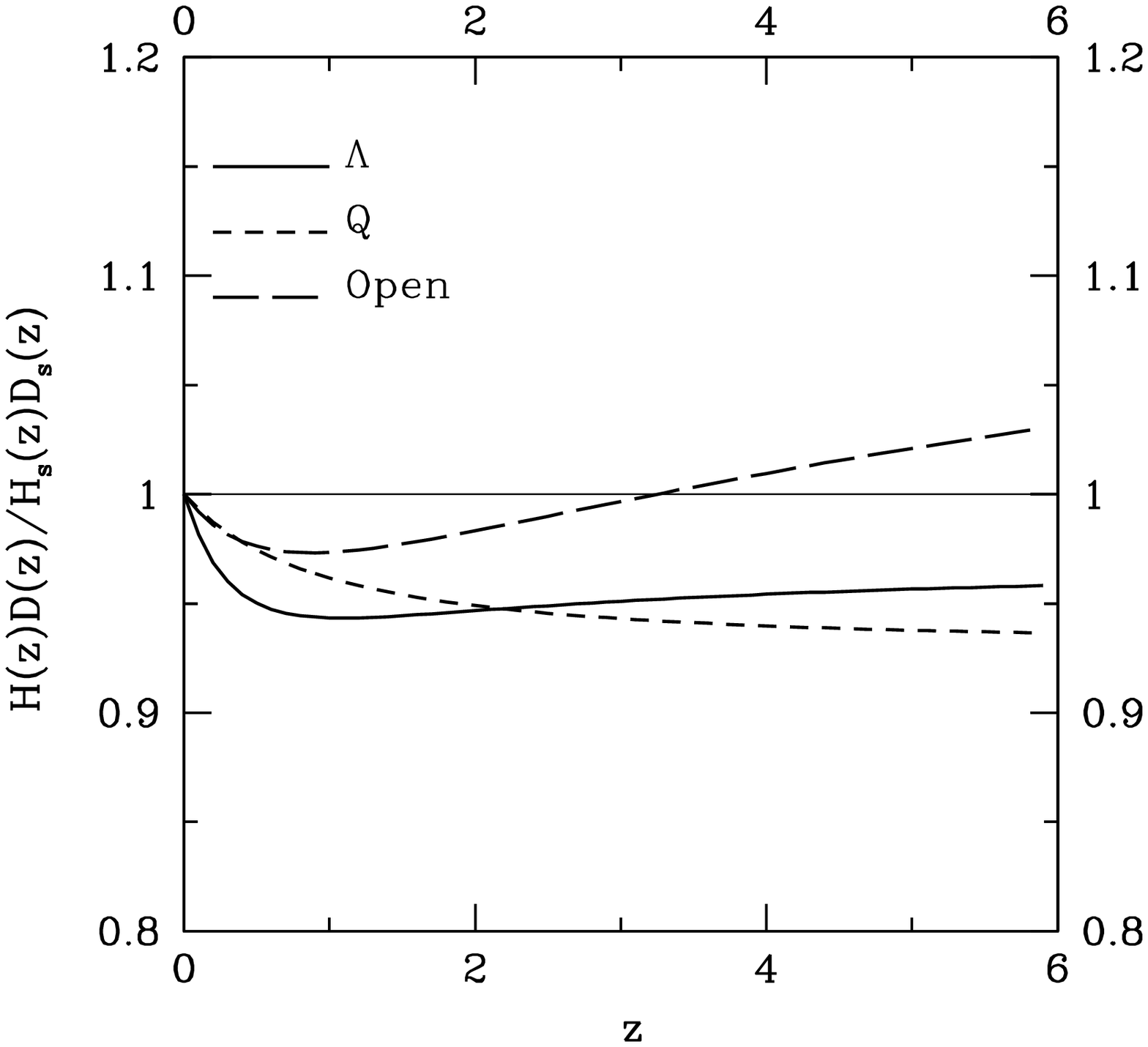}
}
}
\vskip -40pt
\caption{  Ratio of $H(z)\,D(z)$ in a Model to its value in the fiducial
$\Omega_{0}=1$ Model. $\Omega_{0}=0.7$ for all three Models.
}
\end{figure}
\vfill\eject
\newpage

\begin{figure}
\centerline{
\hbox{
\epsfxsize=4.4truein
\epsfbox[55 32 525 706]{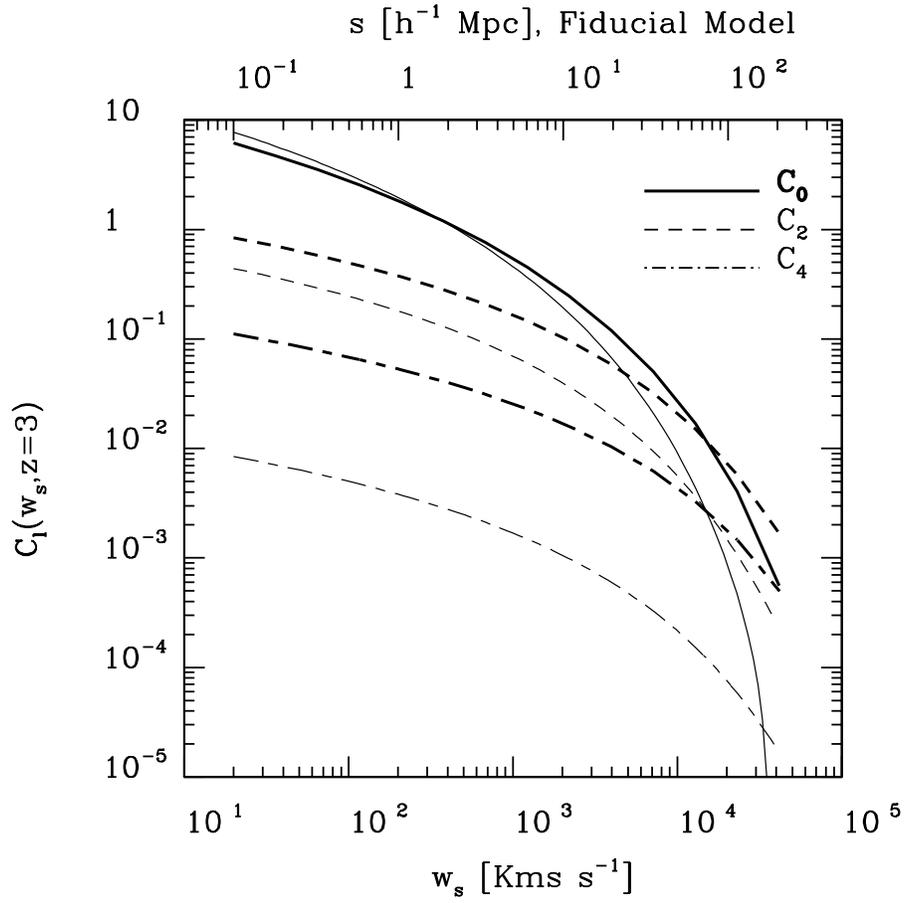}
}
}
\vskip -40pt
\caption{ The light lines corresponding to the fiducial $\Omega_{0}=1$ 
Model. The bold lines correspond to $\Omega_{0}=0.3$,
 $\Lambda_{0}=0.7$ Model. Note that it is the absolute values of the 
coefficients that have been plotted. In particular the quadrupole is 
negative.
}
\end{figure}
\vfill\eject
\newpage

\begin{figure}
\centerline{
\hbox{
\epsfxsize=4.4truein
\epsfbox[55 32 525 706]{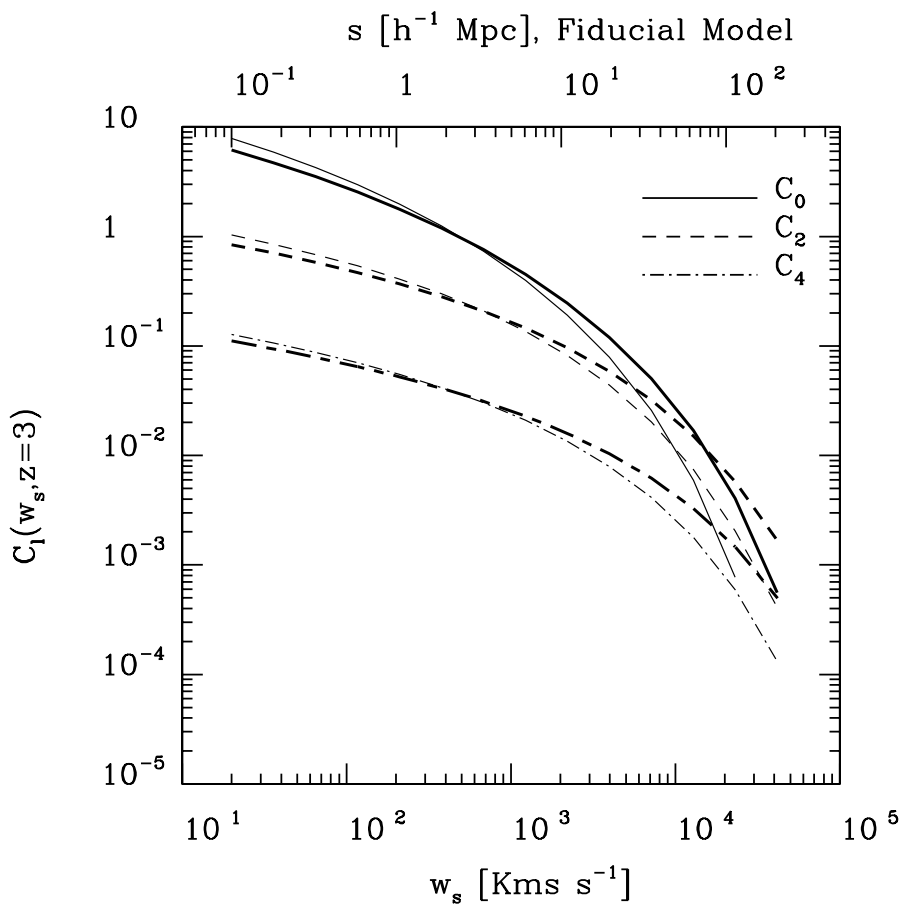}
}
}
\vskip -40pt
\caption{ Both the light and bold lines assume a cosmological geometry 
corresponding to the $\Omega_{0}=0.3$, $\Lambda_{0}=0.7$ Model. The light 
lines correspond to $\Omega_{0}=1$ CDM power spectrum in units of inverse 
velocity. The bold lines correspond to $\Omega_{0}=0.3$,$\Lambda_{0}=0.7$
CDM power spectrum. 
}
\end{figure}
\vfill\eject
\newpage

\begin{figure}
\centerline{
\hbox{
\epsfxsize=4.4truein
\epsfbox[55 32 525 706]{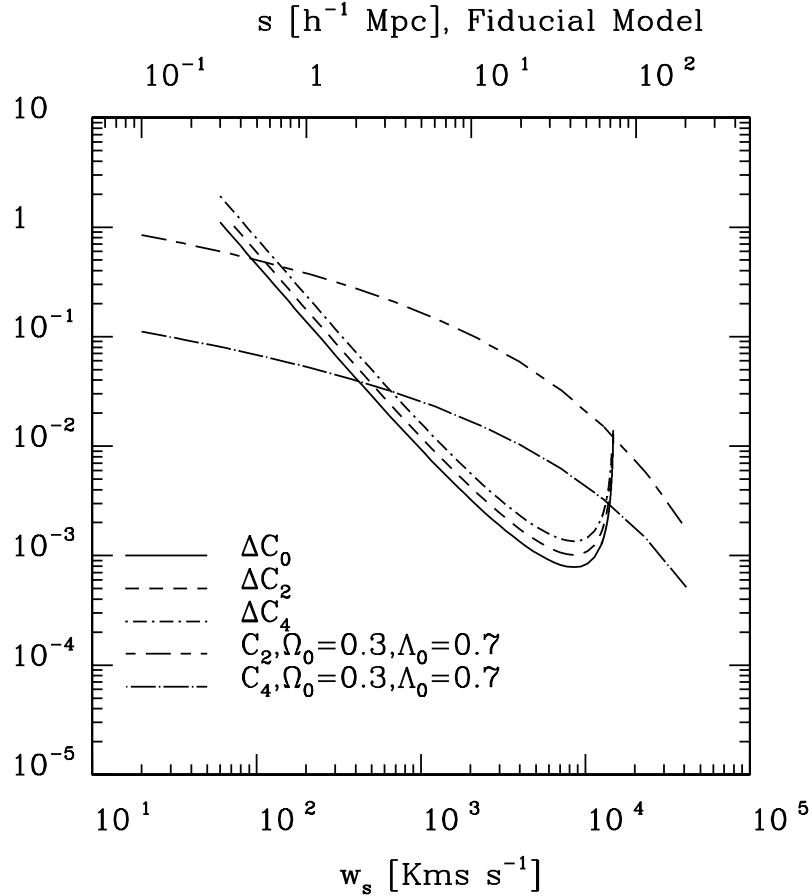}
}
}
\vskip -40pt
\caption{ Figure shows the estimated shot noise error on the the $C_{l}$ coefficients.
Pairs are binned such that each bin centered 
at $w_{s}$ was assumed to have a bin size given by $w_{s} \pm 1/5 w_{s}$. 
This corresponds to binning
where every successive bin is centered at a value $1.5$ times that of the previous
bin. For small separations the error approximately scales as the inverse of the cube 
of the size of the bin. The number of pairs rapidly reduce as separations approach the 
radius of the survey and hence the shot noise error on scales close to the radius 
of the survey is large.}

\end{figure}
\vfill\eject
\newpage

\begin{figure}
\centerline{
\hbox{
\epsfxsize=4.4truein
\epsfbox[55 32 525 706]{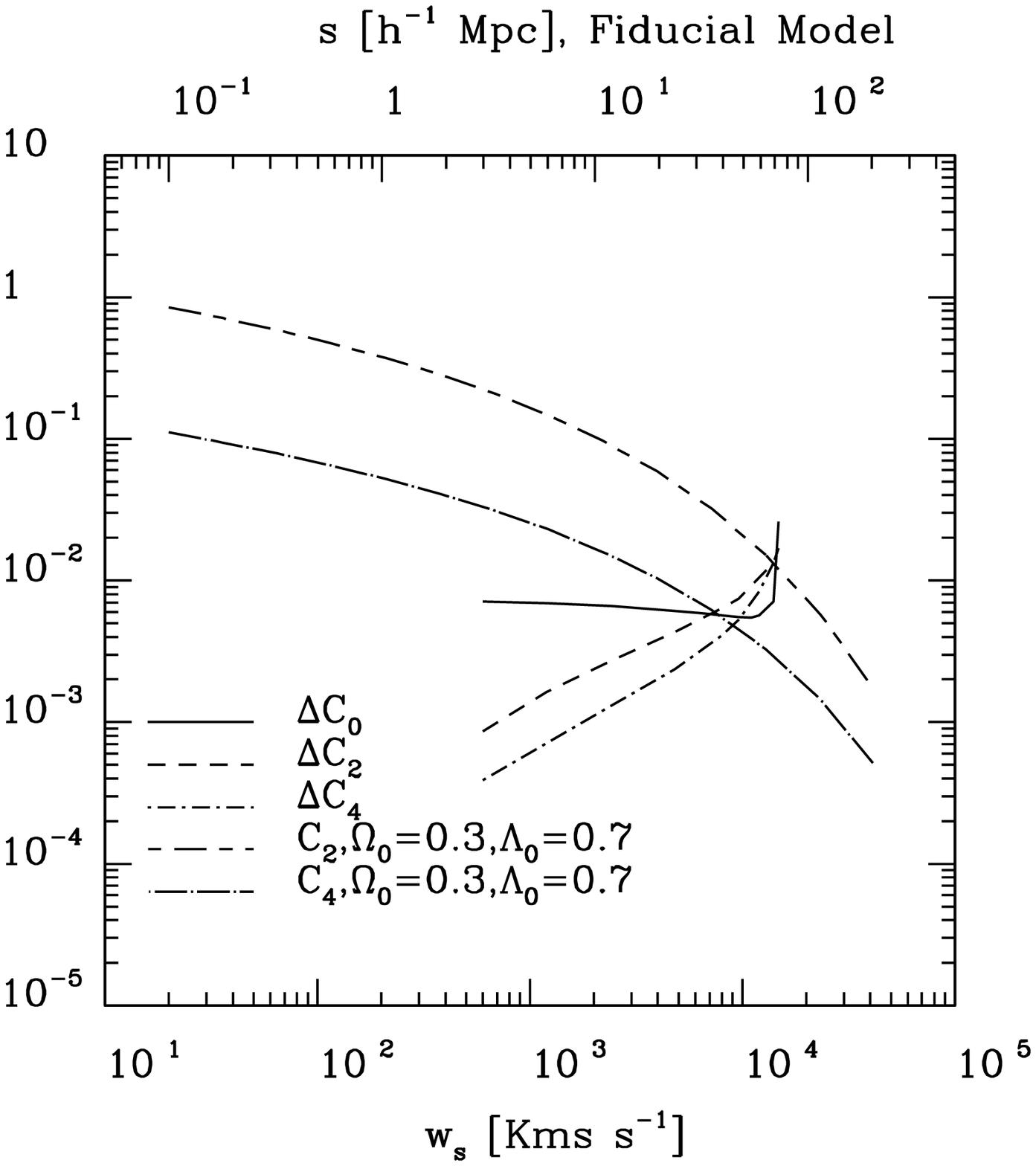}
}
}
\vskip -40pt
\caption{ The cosmic variance error on the $C_{l}$ coefficients are
shown for the fiducial $\Omega_{0}=1$ Model. 
 }
\end{figure}
\vfill\eject
\newpage


\newpage
\begin{thebibliography}{}
\bibitem[]{} Adelberger, K. L., \etal 1998, ApJ, 505, 18
\bibitem[]{} Alcock, C., \& Paczy\'nski, B. 1979, Nature, 281, 4
\bibitem[]{} Babul, A., \& White, S. D. M. 1991, MNRAS, 253, L31
\bibitem[]{} Ballinger, W.E., Peacock, J.A., \& Heavens, A. F. 1996, MNRAS, 282, 877
\bibitem[]{} Bower, R. G., \etal 1993, ApJ, 405, 403
\bibitem[]{} Caldwell, R. R., Dave, R., \& Steinhardt, P. J. 1998, PRL Vol. 80, No. 8
\bibitem[]{} Coles, P., \etal 1998, MNRAS, 300, 183
\bibitem[]{} Dekel, A., \& Lahav, O. 1998, astro-ph/9806193
\bibitem[]{} Eke, V. R., \etal 1996, MNRAS, 282, 263
\bibitem[]{} Fisher, K. B. 1995, ApJ, 448,494
\bibitem[]{} Fisher, K. B., \etal 1994, MNRAS, 267, 927
\bibitem[]{} Garnavich, P., \etal 1998, ApJ, 493, L53
\bibitem[]{} Giavalisco, M., \etal 1998, ApJ, 503, 543
\bibitem[]{} Gazta\~naga, E., \& Baugh, C. M. 1998, MNRAS, 294, 229
\bibitem[]{} Guhathakurta, P., \etal 1990, ApJ, 357, L9
\bibitem[]{} Hamilton, A. J. S. 1992, ApJ, 385, L5
\bibitem[]{} Hamilton, A. J. S. 1993, ApJ, 417, 19
\bibitem[]{} Kaiser, N. 1984, ApJ, 284, L9 
\bibitem[]{} Kaiser, N. 1987, MNRAS, 227, 1
\bibitem[]{} Kodama,,.  H. \& Sasaki, M. 1984, Prog. Theo. Phys. Suppl., 78, 166
\bibitem[]{} Matsubara, T., \& Suto, Y. 1996, ApJ, 470, L1
\bibitem[]{} McGill, C. 1990, MNRAS, 242, 428
\bibitem[]{} Mo, H. J., \& White, S. D. M  1996, MNRAS, 282, 1096
\bibitem[]{} Peebles, P. J. E., \& Ratra, B. 1988, ApJ, 325, L17
\bibitem[]{} Perlmutter, S., \etal 1997, ApJ, 483, 565
\bibitem[]{} Reiss, A. G., \etal 1998, AJ, 116, 1009
\bibitem[]{} Scherrer, R. J., \& Weinberg, D. H. 1998, ApJ, 504, 607 
\bibitem[]{} Steidel, C. C., \& Hamilton, D. 1993, AJ, 105, 2017
\bibitem[]{} Steidel, C. C., \etal 1996, ApJ, 462, L17
\bibitem[]{} Steidel, C. C., \etal 1998, ApJ, 492, 428
\bibitem[]{} Tegmark, M., \& Peebles, P. J. E. 1998, ApJ, 500, L79  
\bibitem[]{} Wechsler, R. H.,  \etal 1998, ApJ, 506, 19
\end{thebibliography}
\end{document}